\documentclass{article}[11pt]

\usepackage{amsmath}
\usepackage{graphicx}
\usepackage{natbib}
\usepackage{theorem}
\usepackage{setspace}
\usepackage{amssymb}
\usepackage{color}
\usepackage{lineno}
\usepackage[margin=2cm]{geometry}

\newcommand{\V}{{\mathsf{FP}}}

\newcommand{\E}{{\mathbb{E}}}
\newcommand{\Pro}{{\mathbb{P}}}

\newcommand{\Rj}{{\mathcal{R}}}
\newcommand{\fwer}{\mbox{FWER}}

\newcommand{\fdr}{{\mbox{FDR}}}

\newcommand{\fdp}{{\mbox{FDP}}}

\newcommand{\paren}[1]{\left(#1\right)}

\newcommand{\st}{\;|\;}

\newcommand{\rbar}{\overline r}
\newcommand{\cbar}{\overline c}

\newcommand{\sev}{\mbox{SEV}}

\newcommand{\SU}{\mbox{SU}}

\begin{document}
\title
{Two step multiple comparisons procedures for positively dependent data with application to detecting differences in human brain network topologies}
\author{Djalel Eddine Meskaldji\footnote{Signal Processing Laboratory (LTS5), Ecole Polytechnique F\'{e}d\'{e}rale de Lausanne (EPFL), Lausanne, Switzerland. Email: djalel.meskaldji@epfl.ch.} \and Patric Hagmann \footnote{Departement of Radiology, University Hospital Center and University of Lausanne (CHUV-UNIL), Lausanne, Switzerland}
\and Jean-Philippe Thiran* \and Stephan Morgenthaler\footnote{FSB/MATHA, Ecole Polytechnique F\'{e}d\'{e}rale de Lausanne (EPFL), Lausanne, Switzerland.}}
\maketitle
\doublespacing
\begin{abstract}
We consider the problem of testing positively dependent multiple hypotheses assuming that a prior information about the dependence structure is available. We propose two-step multiple comparisons procedures that exploit the prior information of the dependence structure, without relying on strong assumptions. In the first step, we group the tests into subsets where tests are supposed to be positively dependent and in each of which we compute the standardized mean of the test scores. Given the subset mean scores or equivalently the subsets p-values, we apply a first screening at a predefined threshold, which results in two types of subsets. Based on this typing, the original single test p-values are modified such that they can be used in conjunction with any multiple comparison procedure. We show by means of different simulation that power is gained with the proposed two-step methods, and compare it with traditional multiple comparison procedures. As an illustration, our method is applied on real data comparing topological differences between two groups of human brain networks.
\end{abstract}


\section{Introduction}
The study of brain connectivity has become an important aspect of neuroscience. It helps to understand brain organization and function \citep{SpornsBook}. Through recent innovations in medical imaging and image analysis, the determination of the interregional brain connectivity became feasible. Brain connectivity is derived either from morphological diffusion or functional neuroimaging data \citep[e.g.,][]{CMTKpaper, Cammoun2012, Friston11, Hagmann2010MR, VandenHeuvel10, Achard2006, Liu08} and is represented by a network (in the graph theory sense) or equivalently by a connection matrix (adjacency matrix) where each cell represents a measure of connectivity between two regions of interests (ROIs) of the brain.
Investigating differences in connectivity between distinct groups of individuals based on connectivity matrices is attractive but is also linked to certain number of problems, among them, the multiplicity of the tests \citep{Meskaldji2013NeuroImageReview}.\\ 


When the comparison between brain networks are studied at the level of nodes (vertices) that represent brain ROIs or connections (edges) linking brain ROIs, a huge number of tests has to be performed on the same data. If the multiplicity is ignored, the risque of committing false discoveries increases. As a consequence, erroneous conclusions are frequently drawn.\\

Consider a set of $M$ hypotheses $h_j, j\in J = \{1, \ldots,M\}$ to be tested. Each hypothesis refers to a single unit called \emph{atom}. An atom could be a pixel, a voxel or a brain region of interest (ROI) in the bio-imaging context, and could be a connection (edge) or node (vertex) in the brain connectivity context.\\

Let $h_j =0$ if the hypothesis regarding the atom, is true and $h_j=1$ otherwise. Let $J^N =\{j\in J: h_{j}=0\}$ the set of true null hypotheses and $J^A=J \backslash J^N$ the set of false null (alternative) hypotheses. The hypotheses are associated to the test statistics $\{Z_1,\ldots,Z_M\}$ or equivalently to the p-values $\{p_1,\ldots,p_M\}$.\\

A multiple testing procedure (MTP) is a decision function $\mathcal{D}: [0,1]^M \rightarrow \{0;1\}^M$ that associate to each set of p-values $\mathcal{P}=\{p_1,\ldots,p_M\}$ a vector $\{R_1,\ldots,R_M\}$ where $R_j =1$ if the hypothesis $h_j$ is rejected and $0$ otherwise. A {\em rejection set} $\Rj$ is a subset of $J$ that indicates the rejected hypotheses, that is, ${\Rj}=\{j \in J: R_j=1\}.$ The number of erroneously rejected hypotheses is the number of false positives $\V =|\mathcal{R} \cap J^N|$.
For example, the non-multiplicity correction procedure (NMCP) consists in rejecting $h_j$ if $p_j$ is less than a predefined level $\alpha$, that is, $R_j =1$ if $p_j\leq\alpha$, and hence ${\mathcal{R}}=\{ j \in J : p_j\leq\alpha\}.$

The probability of committing at least one false rejection is called \emph{the family wise error rate ($\fwer$)}
\begin{equation}\label{example}
\fwer=\Pro(\V>0).
\end{equation}
If all hypotheses are true (i.e., $M_0\equiv|J^N|=M$) and the test statistics are independent, the $\fwer$ of the NMCP is $1-(1-\alpha)^{M}.$ For example, if $M=1000$, and a typical value of  $\alpha =0.05$ is used, the this probability is $\approx 1$. The expected number of false positives is $\E(\V)=\alpha M=50$. Even though these considerations make it clear that a correction for multiplicity should be mandatory, a large number of claims are published without a proper control \citep{BenjaminiSI2010}.\\

Traditional MTPs control the $\fwer$, that is, they guarantee $\fwer \leq \alpha$. A famous such MTP is the \cite{bonferroni1936} procedure which defines the rejection set by ${\Rj}=\{ j \in J : p_j\leq\alpha/M\}.$ In this,

$$\Pro(\V\geq 1)\leq \E\left(\frac{\V}{1}\right)=\sum_{j \in J^N} I\{p_j \leq \frac{\alpha}{M}\}\leq M_0 \frac{\alpha}{M}\leq \alpha,$$
where the first inequality is a particular case of the Markov's inequality.\\

The Bonferroni procedure is the simplest and the strongest procedure in terms of control of $\V=|\Rj \cap J^N|$. It is also a single step procedure, in the sense that each p-value is considered independently of the others.\\

Many other MTPs that control the $\fwer$ have been proposed, although they typically give only a slight improvement over the Bonferroni procedure. This includes step-wise procedures. With this type of procedures, the p-values are considered simultaneously in the sense that a value of each p-value has an in influence on the global inference. Examples of step-wise procedures include \cite{Holm1979}, \cite{SIMES1986}, \cite{HOCHBERG1988}. \\

Detecting real effects is also of great importance in multiple testing. \cite{BenjaminiFDR1995} introduced the false discovery rate ($\fdr$) as an alternative to the $\fwer$ with the aim of increasing the power of detecting true alternatives. The false discovery rate ($\fdr$) is $\fdr = \E (\fdp)$ where the ($\fdp$) is
\begin{equation}
\fdp = \frac{\mathrm{ number\  of\  false\ rejections}}{\mathrm{ number\  of\  rejections}} =
\frac{| {\Rj} \cap J^N |}{| {\Rj}|},
\end{equation}
and the ratio is defined to be 0 if the denominator is 0. Benjamini and Hochberg (1995) proved that $\fdr \leq \alpha$ if ${\Rj} = \{ j:\ p_{(j)} \leq p_{(U)}\}$ where $U = \max\{ j:\ p_{(j)} \leq j\alpha/M\}$. We call this step-up procedure the linear step-up (LSU) procedure from now on.\\

A more general concept of error rates has been recently proposed by \cite{MeskCER2011} and it is called the Scaled Multiple Testing procedures. They showed for example, that if the critical values of the LSU $u_j = \alpha j/M$, are replaced by $\alpha g(j)/M$, where $g(j)$ is a non-decreasing function, the resulting step-up (SU) procedure controls the Scaled Expected Value defined by
\begin{equation}
\mathrm{SEV}=\E\left\{\frac{| {\Rj} \cap J^N |}{g(| {\Rj}|\vee 1)}\right\}.
\end{equation}
This new family of procedures gives a more flexible choice of error rates for the researchers.\\

Among the multiple comparisons procedures that have been proposed in the literature, few of them exploit the dependency structure of the data. The positive dependence that could be present between the test statistics should however be exploited in the aim of increasing the power of detecting true alternatives.\\

In order to face the multiplicity challenge in the presence of positively correlated test statistics, many strategies have been adopted. The first strategy consists in reducing the number of hypotheses by grouping them into subsets in which test statistics are supposed to be positively correlated. The main advantages of dividing the global set of tests into subsets, as has been shown in \cite{BH07,Meskaldji2011},  are (1) the reduction of the number of tests and (2) the reduction of the noise. Both lead to an increase of the power of comparisons at the subsets level.\\

Grouping relies on Tukey's concept called {\em borrowing strength}. This was adopted by \cite{Brilinger1990} in a geographical application where groups were defined on the basis of geographical regions. It was also adopted by \cite{PennyFriston2003} to detect activations in fMRI data. Based on the idea that voxels of a neurological type belonging to a unique anatomical region will usually exhibit positively correlated behavior \citep{PennyFriston2003, Genovese1999functionl}.
In general, methods that use grouping rely on the random field theory, where data is supposed to be smooth and follow a multi-dimensional Gaussian distribution. For this reason, a smoothing has to be applied on the data \citep{NicholsFWERreview2003}. A permutation approach is performed to define active clusters.\\

The same concept has been followed to derive specific statistical methods in the brain connectivity context. \cite{ZaleskyNBS} proposed the network based statistic (NBS) as a method to correct for the $\fwer$, in the framework of multiple comparisons applied to the brain network connections. The method relies on a first identification of connected components (in a graph theory sense), by extra-thresholding the set of p-values $\cal P$ at an arbitrary threshold. An iterative procedure based on permutation testing allows thereafter identifying connected components which carry a between-groups effect.\\

\cite{BH07} proposed a quite different grouping strategy where the choice of groups is defined beforehand using prior information. The advantage of this strategy is that it exploits the positive dependence without relying on the strong assumption of smooth continuous data. Each subset is summarized by the standardized mean of the test scores to derive p-values that are specific to each subset. A weighted \citep{BH97weightedMTP} version of the LSU procedure is applied on these p-values to control the $\fdr$ at the subset level. By specifying non negative weights $w_{j}, j\in J$, \cite{Genovese2006WFDR} showed that if the Bonferroni procedure is applied to $\frac{p_{j}}{w_{j}}$ instead of $p_{j}$, the FWER is controlled as long as $M^{-1}\sum_j w_{j} =1$. \cite{Genovese2006WFDR} also showed that the FDR control procedures benefit by weighting. They showed that if the LSU procedure is applied to the p-values $p_{j}/w_{j}$ instead of $p_{j}$ $\forall j \in J$, the $\fdr \leq \alpha$ as long as $m^{-1}\sum_j {w_{j}} =1$ as before. See also \cite{Roquain2009Wighted}. In \cite{BH07}, the weights are proportional to the size of each subset.\\

It is commonly admitted that most mental diseases or cognitive trait exhibit changes not in the entire brain uniformly but rather specific in functional systems or brain regions and this in different ways and extent. Under this assumption, \cite{Meskaldji2011} proposed a grouped strategy where groups are subnetworks of the global brain network. Each subnetwork is summarized by a meaningful measure (univariate or multivariate) to derive a subnetwork p-value. A multiple comparison procedure is performed to control a desired error rate at the subnetworks level.\\

The question that we will study in this paper is to go beyond the subsets level and investigate the differences at the single nodes/connections by exploiting the positive dependence. \cite{BH07} proposed a two-step method that exploits the positive dependence between tests and showed a great improvement in terms of power. Their method is a two-step method.  The first screening step is the grouping strategy described above. A second step follows by performing an $\fdr$ control procedure on the conditional p-values that belong to the subsets that passed the screening step. The method needs the estimation of the unknown parameters of the problem on which the false positives control is based.\\

We propose a simpler and at the same time a more general approach. We introduce two-step procedures where we do not need the estimation of the conditional p-values. First, we group the family of tests into subsets and we apply a first screening or subset classification at a predefined threshold using the standardized mean of the test scores as a summary statistic for each subset. The first step results in two classes: positive subsets and negative subsets. Based on the first step results, we modify the original p-values such that the modified p-values can be used with any multiple comparison procedure proposed in the literature to control the desired error rate. The modification consists in dividing the p-values belonging to the positive (negative) subsets by a unique coefficient called the \emph{relaxation (tightening) coefficient}.\\ 

The report is organized as follows. We give a general formulation of the 2-step methods and some related results. We next propose methods to estimate the relaxation/tightening coefficient and we show based on various simulation scenarios the benefit of using the proposed methods. Finally we present a practical application on real data, which consists on comparing human brain network topologies across populations.\\
\section{General Formulation and notation}
Based on prior information, we group the $M$ tests into $m$ subsets $ J_1,\ldots, J_m$ such that $\bigcup_{i=1} ^{m}  J_i=J$ and $| J_i|=s_i, i=1,\ldots,m$. The hypotheses $h_{ji}$ are indexed by two indices: the subset index $i$ and the atom index $j$. The hypothesis $h_{ji}$ is associated to a score test $Z_{ji}$. 


When comparing two populations of sizes $n_1$ and $n_2$ respectively, the original data are often of the form
\begin{equation}
\label{dataMatrix}
\mathbb{M}=\left(\begin{array}{cccccccccc}
1&\ldots&1&\ldots&m&\ldots&m\\
X_{111}&\ldots & X_{1j1}& \ldots & X_{1j'm}& \ldots & X_{1Mm}\\
X_{211}&\ldots & X_{2j1}& \ldots & X_{2j'm}& \ldots & X_{2Mm}\\
\vdots &\ldots & \vdots & \ldots & \vdots    & \ldots & \vdots \\
X_{n_1 11}&\ldots & X_{n_1j1} & \ldots & X_{n_1j'm}& \ldots & X_{n_1Mm}\\
Y_{111}&\ldots & Y_{1j1}& \ldots & Y_{1j'm}& \ldots & Y_{1Mm}\\
Y_{211}&\ldots & Y_{2j1}& \ldots & Y_{2j'm}& \ldots & Y_{2Mm}\\
\vdots &\ldots & \vdots & \ldots & \vdots    & \ldots & \vdots \\
Y_{n_2 11}&\ldots & Y_{n_2 j1}& \ldots & Y_{n_2 j'm}& \ldots & Y_{n_2 Mm}
\end{array}\right)
\end{equation}

where the $j^{\mathrm th}$ test statistic $Z_{ji}$ is based on the $j^{\mathrm
  th}$ column of $\mathbb{M}$.  Usually, $Z_{ji}$ is 
approximately (or exactly) $\mathcal N(\mu_{ji},1)$.\\

The p-values associated with the tests are $p_{ji}, j\in J$ where $p_{ji} = \overline{\Phi}(Z_{ji})$, $\overline{\Phi} = 1- \Phi$ and $\Phi$ denotes the standard normal cumulative distribution function.
Without loss of generality, we consider one-sided tests $h_{ji} =1$ if $ \mu_{ji}>0$ and $h_{ji}=0$ otherwise.\\

As in \cite{BH07} the following mixture effect model is considered. \emph{Null subsets} are subsets that only contain null hypotheses. Otherwise, the subset is called \emph{affected}. Let $I_0=\{i:1,\ldots,m| \sum_{j \in J_i}h_{ji}=0\}$ and $I_1=\{i:1,\ldots,m| \sum_{j \in J_i}h_{ji}>0\}$ the indices of null subsets and affected subsets respectively. The number of null subsets is $|I_0|=m_0$ and the number of affected subsets is $|I_1|=m_1$.
The proportion of non-null hypotheses in the subset $J_i$ is $\pi_i=\frac{1}{s_i}\sum_{j \in J_i}h_{ji}$.\\
%
%
The information of each subset is summarized by a summary statistic, the standardized mean $T_i=\sum_{j\in J_i} Z_{ji} /\sigma_i$ where $\sigma_i$ is the standard deviation of the subset tests $Z_{ji}: j\in J_i$. To model the dependence inside each subset $J_i$, let $\rho_{ji}= corr(T_i;Z_{ji})$ be the correlation between each test $Z_{ji} \in J_i$ with its corresponding subset standardized mean $T_i\sim \mathcal N(\mu_i,1)$. We assume that $ \rho _{ji} >\frac{1}{\sqrt{s_{i}}}$, which corresponds to $corr(Z_{ji},Z_{li})>0\mbox{ for } (j,l)\in J_i ^2, \; i \in I_1$ and $ \in j\neq l.$ We also assume that $\rho_{ji}=\frac{1}{\sqrt{s_i}}$ for $(j,l)\in J_i ^2, \; i \in I_0.$ and $j \neq l$.\\

According to this model, we have the following distributions.
\begin{equation}
\left( T_{i}|Z_{ji}=z\right) \sim \mathcal{N}\left( \rho_{ji}z,\left(
1-\rho _{ji}^{2}\right) \right)=\mathcal{N}\paren{\frac{1}{\sqrt{s_i}}z, \paren{1-\frac{1}{s_i}}}.
\end{equation}
for $j\in J^N$ and $i \in I_0$.
\begin{equation}
\left( T_{i}|Z_{ji}=z\right) \sim \mathcal{N}\left( \mu _{i}+\rho
_{ji}z,\left( 1-\rho _{ji}^{2}\right) \right)
\end{equation}
for $j\in J^N$ and $i \in I_1$.\\

Suppose that the p-values corresponding to the summary statistics are $P_i (i=1,\ldots,m)$, that is, $P_i=\overline\Phi(T_i)$. 
The central limit theorem allows us to make this approximation especially when the size $s_i$ becomes quite large.\\

Let $P_{(1)} \leq \cdots \leq P_{(m)}$ denote the sorted subsets' p-values and let $T_{(1)} \geq \cdots \geq T_{(m)}$ denote the sorted summary statistics. The first step consists in comparing the subset p-values $P_i\; (i=1,\ldots,m)$ to a predefined threshold $U$. This screening results in two classes of subsets. Let $I^+=\{i: P_i \leq U\}$ and $I^-=I\backslash I^+$ be the positive subsets and the negative subsets respectively, and let $M^+=\left|\bigcup_{i\in{I^+}}J_i\right|$ be the number of atoms in positive subsets, and $M^-=M-M^+$ the number of atoms in negative subsets.\\

If the user is interested by the results at the subsets level, the threshold $U$ has to be defined on the basis of an MTP. One could use the the Bonferroni threshold $U=\alpha/m$ to control the $\fwer$ at the subsets level \citep[see][]{Meskaldji2011}.
In \cite{BH07}, the threshold $U$ is defined on the basis of the LSU procedure. This threshold is $U=\max\{ i:\ P_{(i)} \leq i\alpha/m\}$.\\

\cite{BH07} showed by simulations in the $\fdr$ case and \cite{Meskaldji2011} proved analytically in the Bonferroni case that if a subset of size $s_i$ contains more than $\sqrt{s_i}$ false hypotheses, the power of detecting an affected subsets is larger than detecting the contained false hypotheses using the same MCP at the atoms level, that is, for $j \in J_i \cup J^A$,
$$ \pi_i \geq 1/\sqrt{s_i} \Rightarrow \Pro\left(P_i\leq \frac{\alpha}{m}\right)> \Pro\left(p_{ji}\leq \frac{\alpha}{M}\right).$$

We will also consider in this paper, the threshold defined by the NMCP, that is, $U=\alpha$. The use of this threshold will not guarantee the control of $\V$ at the subsets level, but our aim is to maximize the probability of detecting real effects by controlling a metric of $\V$ at the level of atoms and a less stricter screening could be beneficial.\\

Based on the results of the first step, the researcher can perform a MCP at the level of single hypotheses to control the type I error rate of his/her choosing. This can be afforded by computing the test p-values conditioned by the statistical results at the first step \citep[see][]{BH07}. However, this solution is time consuming in large data and the control of the $\V$ is not guaranteed for small samples by this procedure.\\

\section{Relaxation methods}
We propose a procedure where we do not need to estimate the unknown parameters and to compute the conditional p-values. We work directly with the unconditioned (original) p-values. Our proposed two-step procedure works as follows: We divide the original p-values inside the positive (negative) subsets by a positive number called the \emph{relaxation (tightening) coefficient} and then perform a new multiple comparison on the modified p-values.\\

Consider first, the control of the FWER. In the second step, the original p-values will be compared to $r \alpha/M$ and $\rbar \alpha/M$ for the p-values belonging to the positive and the negative subsets respectively, or equivalently, the scores $Z_{ji}$ are compared to $c=\Phi^{-1}(1-r\alpha/M)$ and $\cbar=\Phi^{-1}(1-\rbar\alpha/M).$ 
This is equivalent to performing the Bonferroni procedure on the set of the modified p-values $\{p_{ji}/r \st i \in I^+\} \bigcup \{p_{ji}/\rbar \st i \in I^-\}$.\\

In order to give an advantage to the positive subsets, we suppose that the tightening coefficient $\rbar \in [0,1]$ with the convention that $p_{ji}/\rbar \equiv 1$ if $\rbar=0$. The relaxation coefficient $r$ and the tightening coefficient $\rbar$, depend on the threshold $U$ used in the first step. The control of the $\fwer$ at level $\alpha$ is guaranteed if the relaxation and the tightening coefficients are chosen to satisfy
$$\mbox{PFER}=\E \paren {\V}\leq \alpha.$$
We can prove that the modified p-values $p_{ji}/r \st i \in I^+$ and $p_{ji}/\rbar \st i \in I^-$ that satisfy this FWER control can be used with the LSU or the scaled SU procedures to control the $\fdr$ or the $\sev$, at level $\alpha$. To show this assertion, we use the concept of weighted MTPs. Set $w_{ji}=r$ for all $i \in I^+$ and $w_{ji}=\rbar$ for all $i \in I^-$. Suppose that the PFER is controlled at level $\alpha$, that is, $\E (\V)\leq \alpha$, we have
\begin{eqnarray*}
\sum_{j\in J^N}&& \Pro \paren {\frac{p_{ji}}{w_{ji}}\leq \frac{\alpha}{M}}\\
=&&\sum_{j\in J^N \cap I^+} \Pro \paren {\frac{p_{ji}}{r}\leq \frac{\alpha}{M}}+\sum_{j\in J^N \cap I^-} \Pro \paren {\frac{p_{ji}}{\rbar}\leq \frac{\alpha}{M}}\\
=&&\sum_{j\in J^N \cap I^+} \Pro \paren {p_{ji}\leq \frac{\alpha r}{M}}+\sum_{j\in J^N \cap I^-} \Pro \paren {p_{ji}\leq \frac{\alpha\rbar}{M}}\\
=&&M^+ \frac{\alpha r}{M} +M^- \frac{\alpha \rbar}{M}\leq \alpha.
\end{eqnarray*}
Here, we used the fact that the p-values corresponding to the null hypotheses $h_{ji}, j \in J^N$ are uniformly distributed in the Gaussian case.
The last inequality implies that $\sum_{j\in J^N} w_{ji}= M^+ r + M^- \rbar \leq M$. Thus, the modified p-values could be used, for example, to control the $\fdr$ by the LSU procedure.\\

Let $\Rj_+ =\{j \in \bigcup_{i\in{I^+}}J_i \st p_{ji}/r\leq \alpha/M \}$ and $\Rj_- =\{j \in \bigcup_{i\in{I^-}}J_i \st p_{ji}/\rbar\leq \alpha/M \}$ be the rejected hypotheses by the second step in the positive and the negative subsets respectively. The set of all rejections is $\Rj = \Rj^+ \cup \Rj^-$. After performing the second step, consider these two quantities: the expected number of $\V$ in the positive subsets $\E(|J^N \cap \Rj^+|)$ and the expected number of $\V$ in the negative subsets $\E(|J^N \cap \Rj^-|).$\\

Given $\rho_{ji},m_0,m,s_i,\mu_i, U,c$ and $\cbar$, and that $\E(I^+)=m\Pro\left(P_{i} \leq U \right)$. Therefore
\begin{eqnarray}
\nonumber &\E&(\left|J^N \cap \Rj^+\right|)=\E\left(\E\left(\left|J^N \cap \Rj^+\right||I^+\right)\right)\\
&\nonumber=& \E\left( \sum_{i\in{I^+}}  \sum_{j \in J_i \cap J^N} \Pro_{h_{ji}=0}(Z_{ji}>c|P_{i}\leq U) \right)\\
&\nonumber=& \sum_{i\in{I^+}} s_i (1-\pi_i) \E\left( \frac{\Pro\left(P_{i}\leq U|Z_{ji}=z\right) \Pro_{h_{0}}\left(Z_{ji} \geq z\right)} {\Pro\left(P_{i} \leq U \right)} \right)\\
 &=& \sum_{i\in{I^+}} s_i \int_{c}^\infty \frac{m_0 \overline\Phi\left(C_0\right)+m_1 (1-\pi_i) \overline\Phi\left(C_{\mu_i}\right)}{m_0 U + m_1 \overline\Phi(\Phi^{-1} (1-U )-\mu_i ) } \varphi(z)dz,
\end{eqnarray}
where $\varphi$ is the probability density function of the normal distribution, $C_0=\frac{\Phi^{-1} (1-U)-\rho_{ji} z}{\sqrt{1-\rho_{ji}^2 }}$ and $C_{\mu_i}=\frac{\Phi^{-1}(1-U)-\mu_i-\rho_{ji} z}{\sqrt{1-\rho_{ji}^2 }}$. Similarly, the quantity $\E\left( \mathbb(\V|I^-)\right)$, which is $\E(\left|J^N \cap \Rj^-\right|)=$ is given by
\begin{equation}
\sum_{i\in{I^+}} s_i \int_{\cbar}^\infty \frac{m_0 \Phi\left(C_0\right)+m_1 (1-\pi_i) \Phi\left(C_{\mu_i}\right)}{m_0 (1-U) + m_1 \Phi(\Phi^{-1} (1-U )-\mu_i ) } \varphi(z)dz.
\end{equation}
Consider now, the particular case where $\mu_{ji}=\Delta$ for all $j \in J^A$, $\pi_i =\pi$ and $s_i =s$ for all $i \in I$. The control of the $\fwer$ at level $\alpha$ is guaranteed if the relaxation and the tightening coefficients are chosen to satisfy $$\E \paren {\V}=\E(|J^N \cap \Rj^+|)+\E(|J^N \cap \Rj^-|)\leq \alpha.$$ In order to achieve this goal, the parameter $\rho_{ji}$ is set to the least favorable value of $1/\sqrt{s}$ which corresponds to the case where the atoms inside the same subset are independent. For the parameter $\Delta$ we chose either $\infty$, which corresponds to the more conservative lower bound, or the mean of the $m_1 s$ largest values of the scores.\\

For a fixed value of the tightening coefficient $\rbar$, the relaxation coefficient $r$ is chosen sufficiently small such that the expected number of false positives $\E(\V)\leq \alpha$, for all possible values of $m_0$ and $\pi_i$. This condition guarantee the strong control of the $\fwer$ by Markov's inequality. Note that we have already proved that the modified p-values $p_{ji}/r \st i \in I^+$ and $p_{ji}/\rbar \st i \in I^-$ can be used with the LSU procedure to control the $\fdr$ at level $\alpha$.\\

\section{Simulations}
We compared the performance of different two-step methods with the AWA, by considering different simulation settings whose results are presented here.
\subsection{Simulation of partially affected subsets}
\begin{figure*}[ht!]
  \centering
\makebox{\includegraphics[angle=270,width=\textwidth]{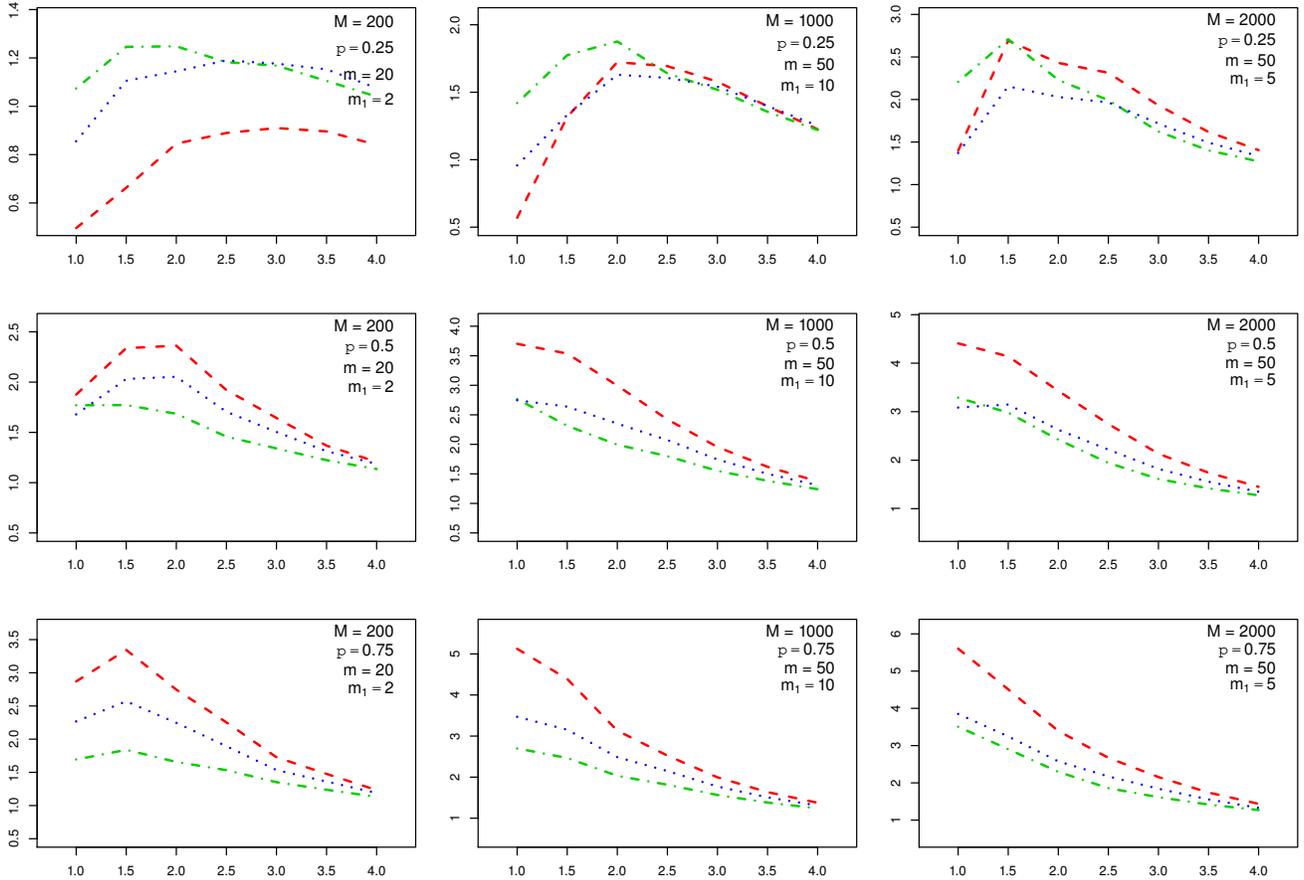}}
  \caption[Ratio of average power of the different relaxed methods over the average power of the AWA against the raw effect, using the Bonferroni procedure to control the FWER in various situations.]{Ratio of average power of the different relaxed methods over the average power of the AWA against the raw effect, using the Bonferroni procedure to control the FWER in various situations. The RMWC (dashed line), the RMNC (dashed-points line) and the RMIO (points). The number of atoms is either $M=100, 500,$ or $1000$. The number of subsets is $m=20, 50,$ or $50$ and $m_1=2,5$ or $10$. In each situation $\pi= 0.25, 0.5$ or $0.75$ as indicated. The number of simulations is 1'000.}\label{BonfSimul}
\end{figure*}

We simulate a set of $M$ hypotheses. We divide the subset of $M$ hypotheses into $m$ subsets with different sizes $s_i, i=1,\ldots,m$ randomly chosen. Among the $m$ subsets, we chose randomly $m_1$ subsets which will contain the effect. In each of the $m_1$ subsets we select randomly a set of hypotheses for which we simulate a test score $Z$ as a random realization of the shifted standard normal distribution with mean $\Delta$, that is, $Z\sim \mathcal{N}(\Delta, 1)$. The average proportion of atoms with effect in the $m_1$ subsets is $\pi$. For all the remaining hypotheses, either in the $m_1$ subsets (containing the effect) or in the remaining $m-m_1$ subsets (without effect), the test scores are random realizations of $\mathcal {N}(0,1)$. The positive dependence is modeled by the proportion of affected atoms in each subset.\\
\begin{figure*}[ht!]
  \centering
\makebox{\includegraphics[angle=270,width=\textwidth]{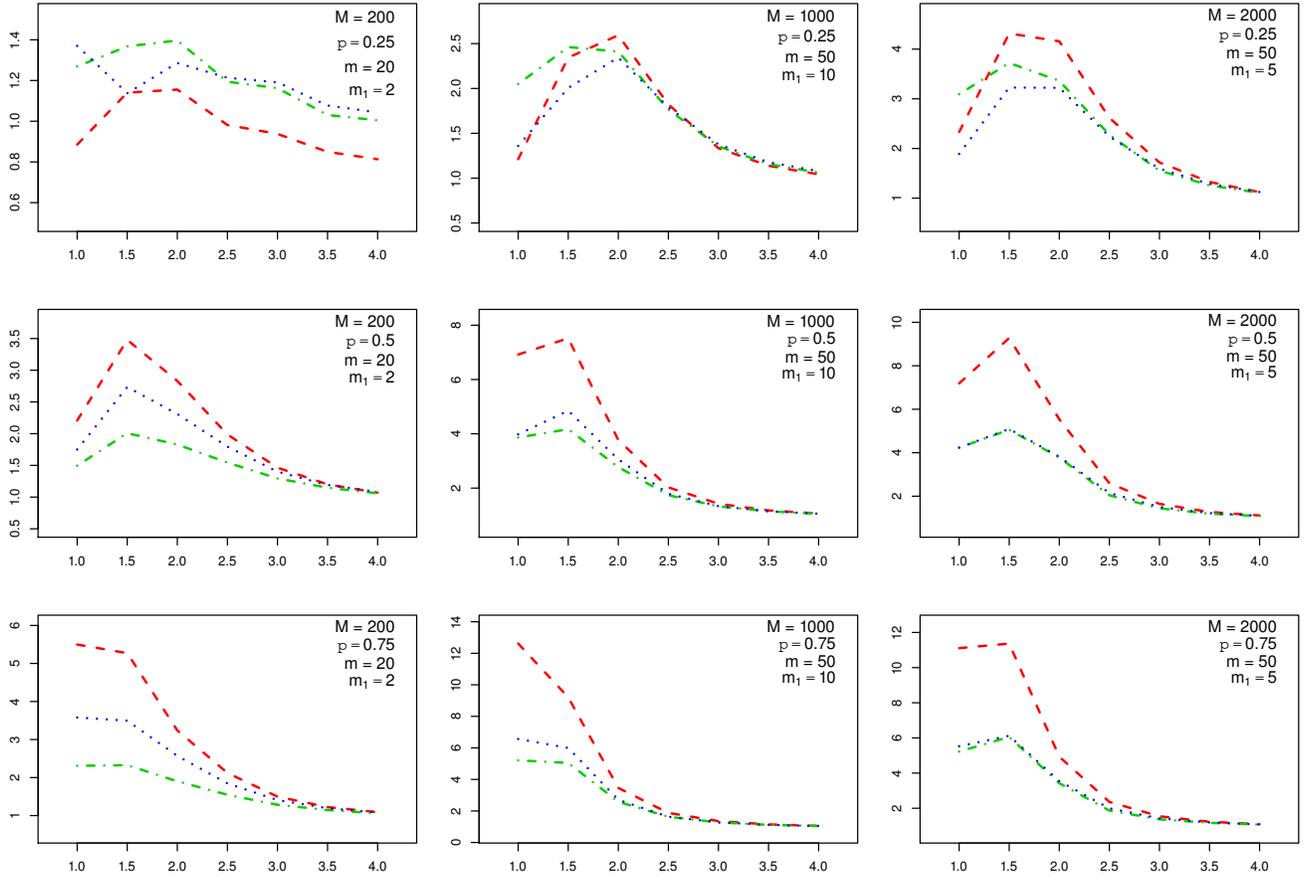}}
  \caption[Ratio of average power of the different relaxed methods over the average power of the AWA against the raw effect, using the LSU procedure to control the FDR in various situations.]{Ratio of average power of the different relaxed methods over the average power of the AWA against the raw effect, using the LSU procedure to control the FDR in various situations. The details are the same as in Figure \ref{BonfSimul}.}\label{BHSimul}
\end{figure*}

We simulated global sets with different sizes $M=200, 1000,$ or $2000$ atoms, with $m=20$ or $50$ subsets and $m_1=2,5$ or $10$ subsets with effect. The average proportion of atoms with effect (within the $m_1$ subsets) was either $\pi= 0.25, 0.5$ or $0.75$. We compared the relaxed method to the AWA in term of average power. Concerning the screening in the first step of the relaxed method we used either a multiplicity correction or no correction. We call these methods Relaxed Method With Correction (RMWC) and Relaxed Method with No Correction (RMNC) respectively. In these two cases, we considered only positive subsets in the second step. We also consider the case where we consider the p-values in the negative subsets. We call this method the Relaxed Method In/Out (RMIO). In this case, the tightening coefficient was set to be 0.5 and we used a multiplicity correction in the first step. See Table~\ref{ParII: Tab: Relaxed Methodes}.\\

\begin{table}[h!]
\centering
\begin{tabular}{|c|c|c|c|}
  \hline  \hline
Name & Step 1 & Step 2 & Consider negative subsets \\  \hline  \hline
RMNC & NMCP & Bonf/LSU & No \\
RMWC & Bonf/LSU & Bonf/LSU & No  \\
RMIO & Bonf/LSU & Bonf/LSU & Yes \\  \hline  \hline
  \hline
\end{tabular}
\caption{Relaxed methods combinations.}
\label{ParII: Tab: Relaxed Methodes}
\end{table}

\begin{table}[h!]
  \centering
\begin{tabular}{|l|c|c|c|c|}
  \hline \hline
   & $\mathrm{AWA}$ & $\mathrm{RMWC}$ & $\mathrm{RMNC}$ & $\mathrm{RMIO}$\\ \hline \hline
  $\mathbb{E}(\V)$ & 0.046 & 0.035 & 0.044& 0.043\\ \hline
  $\mathrm{FDR}$ & 0.057 & 0.053 & 0.055&0.05\\
  \hline \hline \hline
\end{tabular}
\caption[The estimated expected number of false positives, $\mathbb{E}(\V)$ in the Bonferroni case and the estimated FDR in the LSU case.]{The estimated expected number of false positives, $\mathbb{E}(\V)$ in the Bonferroni case and the estimated FDR in the LSU case, for the different methods AWA, RMWC, RMNC and RMIO.}\label{FPsimul}
\end{table}

Figures \ref{BonfSimul} and \ref{BHSimul} show the ratio of the average power of the relaxed methods over the average power of the AWA, in different situations, when using the Bonferroni or the LSU procedures. On the other hand, we reported in Table \ref{FPsimul}, the estimated $\mathbb{E}(\V)$ in the Bonferroni case, and the estimated FDR in the LSU case.\\

The simulations illustrate the gain of power obtained by using the relaxed methods. The relaxed two-step methods perform almost always better than the usual AWA. The gain obtained by the two-step methods reaches more than 5 times the power of the AWA when the raw effect $\Delta$ is small. This corresponds to situations with small real difference or small sample size. The gain is realized even though less false positives are observed. The gain increases as the size of the subsets increases. However, when the proportion of affected atoms in partially affected subsets is small, the relative gain diminishes and may become less then one especially with small subset sizes. When $\Delta$ becomes large, all methods, including the AWA are equivalent.\\
\begin{figure*}[ht!]
  \centering
\makebox{\includegraphics[angle=270,width=\textwidth]{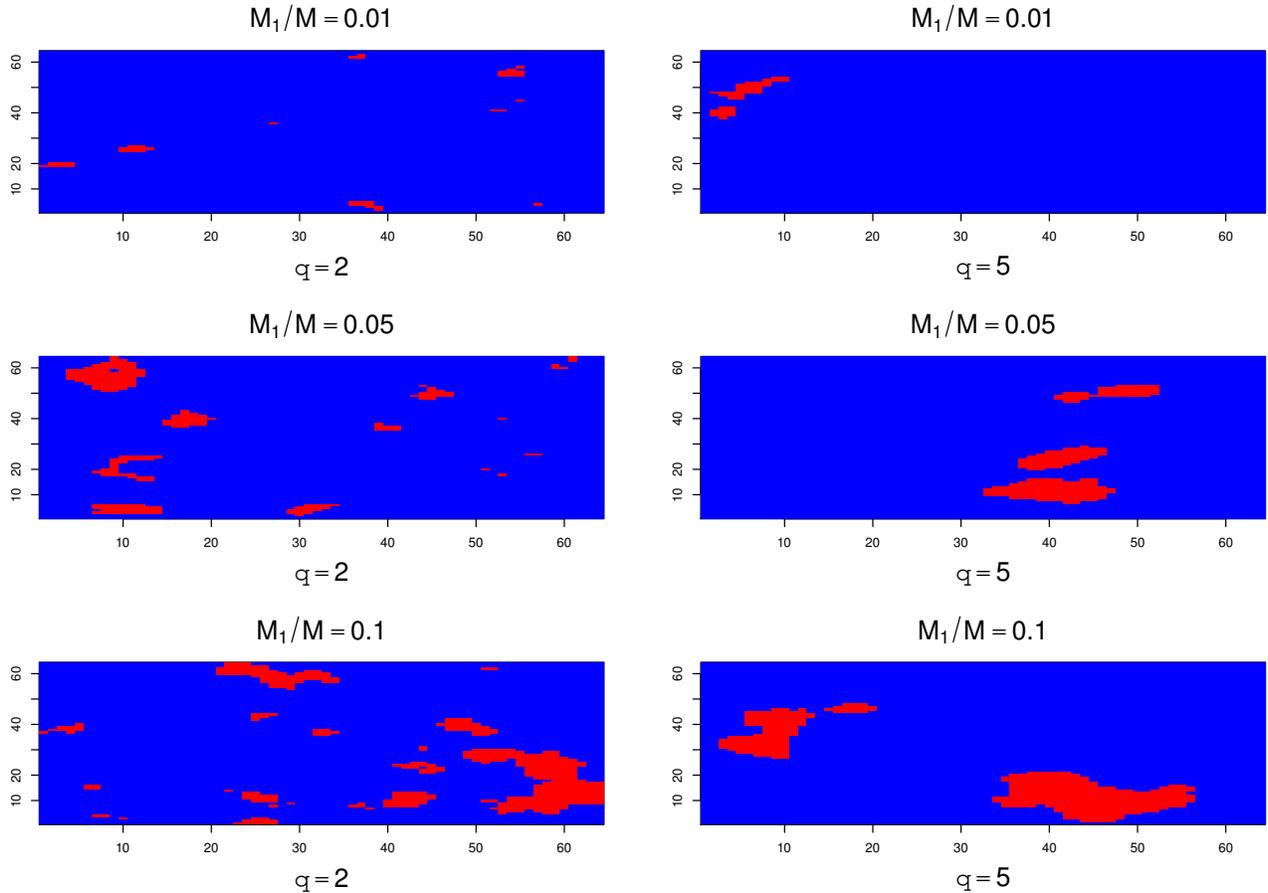}}
\caption[A simulation of the affected atoms in the correlated case (without noise).]{A simulation of the affected atoms in the correlated case (without noise). The red regions represent the positive effect. $M=4096,M_1/M=0.01, 0.05$ or $0.1$, $\theta=2$ or $5$ as indicated.\label{PartII: sim: correlated case snapshots chap 6}}
\end{figure*}

The RMNC is more stable in terms of gain because in this case small proportions are easily detected. The RMWC seems to perform well when the proportion $\pi$ becomes larger which is directly related to the appropriate choice of the decomposition. The RMWC should be chosen when we have more confidence on the network decomposition. Otherwise, RMNC is preferable as it has a less strict screening in the first step. The RMNC and the RMIO have approximately the same performances. Both could be used to detect isolated effects.

\subsection{Simulation of the correlated case}
\begin{figure*}[ht!]
  \centering
\makebox{\includegraphics[angle=270,width=\textwidth]{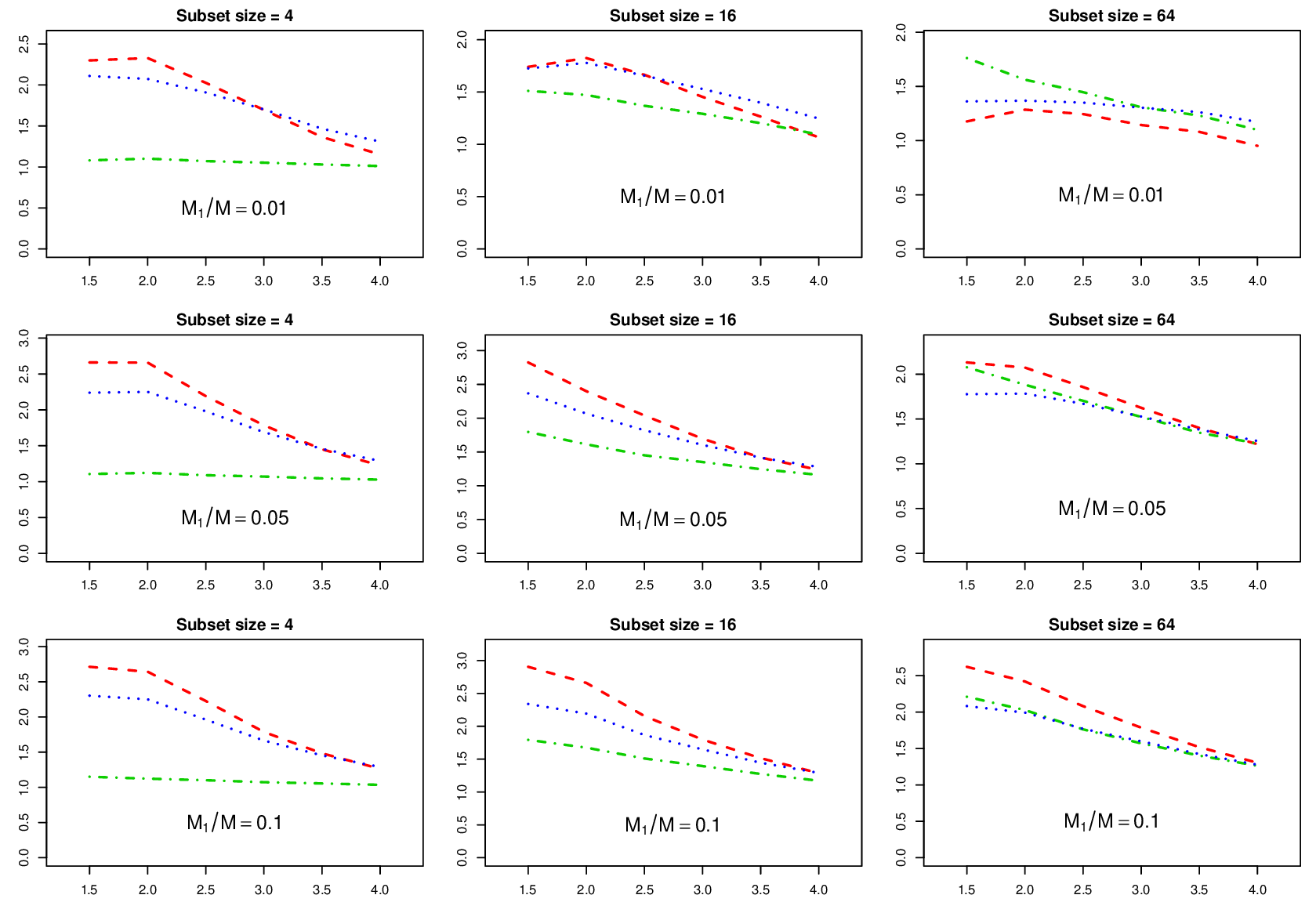}}
\caption[Ratio of average power of the different relaxed methods over the average power of the AWA against the raw effect $\Delta$, using the Bonferroni procedure to control the FWER in different situations.]{Ratio of average power of the different relaxed methods over the average power of the AWA against the raw effect $\Delta$, using the Bonferroni procedure to control the FWER in different situations. The RMWC (dashed line), the RMNC (dashed-points line) and the RMIO (points). The parameter $\theta=2$. The other parameters are $s_i =4, 16\mbox{ or }64$ and $M_1/M=0.01, 0.05\mbox{ or }0.1$ as indicated. \label{PartII: Sim: Correlated case theta = 2}}
\end{figure*}
In the precedent simulations, we compared the relaxed methods and the AWA by considering partially affected subsets. In a different setting, we consider in this section, a model of 2D images, that integrates positive correlations between atoms, not only those contained in partially affected subsets, but a positive correlation between all atoms. We suppose also, in this section, that the researcher has no prior information about the segmentation of the global region of interest. So, the researcher decides to choose the simplest decomposition, that is, the decomposition that consists in dividing the global set into equal square subsets of size $\sqrt{s_i}×\sqrt{s_i}$.\\

The library "fields" of the software "R" affords the simulation of images with correlated atom values. Specifically, the covariance between two pixels/atoms $a$ and $a'$ is proportional to $exp\{(-D/\theta)\}$, where $D$ is the Euclidian distance between $a$ and $a'$ and $\theta$ is a scale parameter.
We generate random images that contain a proportion $\frac{M_1}{M}$ of affected atoms using the "fields" library of the "R" software as flows.
\begin{itemize}
\item We generate a field (image) of size $\sqrt{M}×\sqrt{M}$ using the function "stationary.image.cov".
\item After sorting all the generated values of the image, the $M_1$  largest values are set to $\Delta>0$ and represent the affected atoms. The remaining ones are set to zero.
\item A standard gaussian white noise is added to the resulting image.
\end{itemize}
We presented some snapshots of the correlated case (without noise) in Figure~\ref{PartII: sim: correlated case snapshots chap 6}.\\

\begin{figure*}[ht!]
  \centering
\makebox{\includegraphics[angle=270,width=\textwidth]{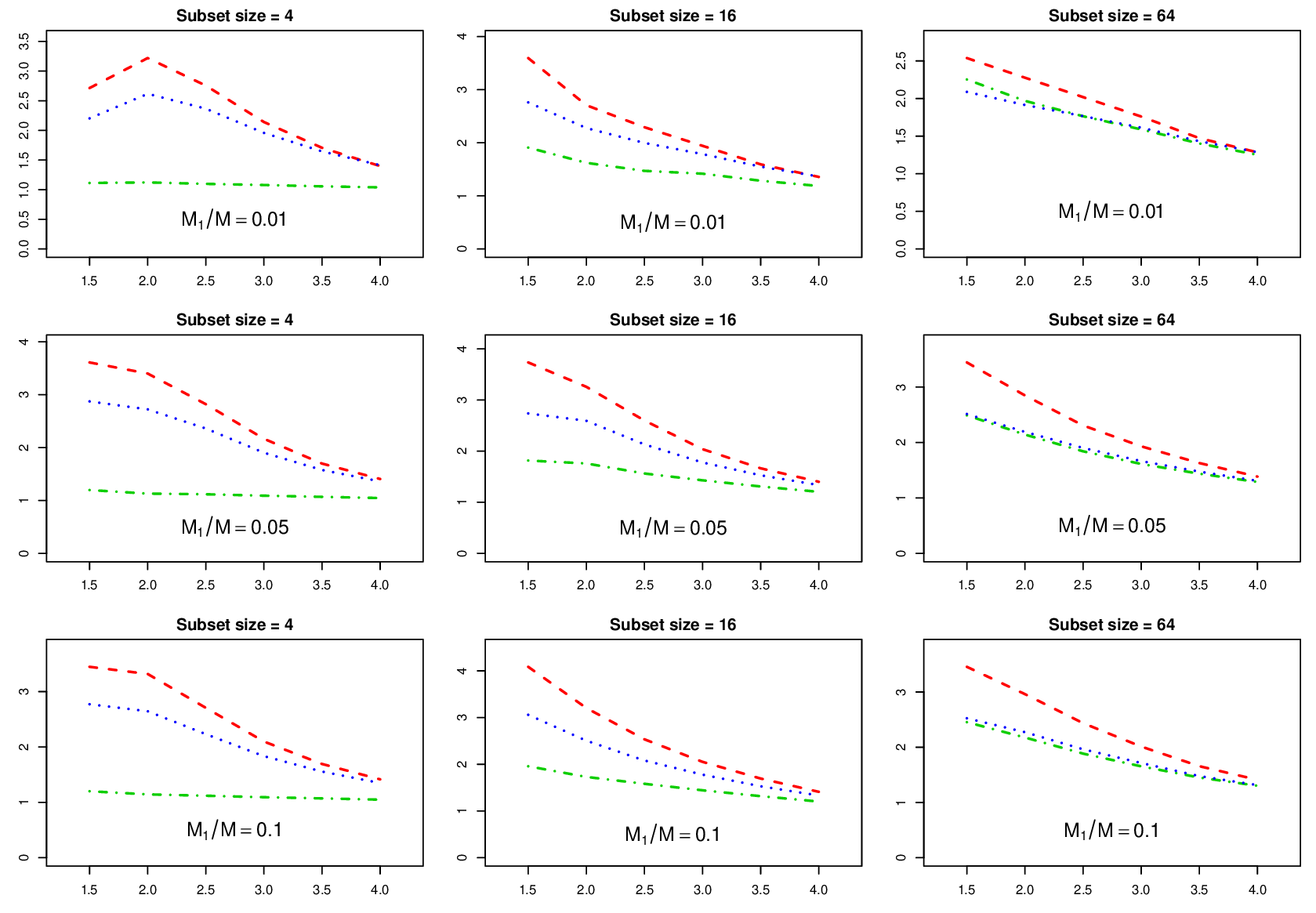}}
\caption[Ratio of average power of the different relaxed methods over the average power of the AWA against the raw effect $\Delta$, using the Bonferroni procedure to control the FWER in different situations.]{Ratio of average power of the different relaxed methods over the average power of the AWA against the raw effect $\Delta$, using the Bonferroni procedure to control the FWER in different situations. The RMWC (dashed line), the RMNC (dashed-points line) and the RMIO (points). The parameter $\theta=5$. The other parameters are $s_i =4, 16\mbox{ or }64$ and $M_1/M=0.01, 0.05\mbox{ or }0.1$ as indicated.\label{PartII: Sim: Correlated case theta = 5}}
\end{figure*}

Figures \ref{PartII: Sim: Correlated case theta = 2} and \ref{PartII: Sim: Correlated case theta = 5} show the ratio of the average power of the relaxed methods over the average power of the AWA, when using the Bonferroni procedure. These two figures show again the potential gain obtained by the relaxed methods. Although the decomposition is not based on prior knowledge, the relaxed methods perform well in the presence of positive correlation.

\section{Application to structural connection matrices of the human brain}
The purpose of this section is to give a real application of the proposed strategies to compare normalized whole-brain structural connection matrices derived from diffusion MRI tractography.
\subsection{Description of the data}
The practical example consists in comparing two groups of brain connectivity matrices on the dataset used in \cite{Hagmann2010} which consists of 30 connection matrices. The connection matrices are derived from an MRI acquisition and well established algorithmic procedures as described in \cite{Hagmann2008,Cammoun2012,CMTKpaper}.\\

We define two groups based on the age of the subjects: 16 pre-school children and 14 adolescent children. {In such connection matrices every network node corresponds to a brain cortical area and every edge corresponds to the white matter structural connectivity between two cortical areas as measured with tractography. The ROIs are chosen on the basis of the Free-Surfer segmentation of the brain cortex.} \\

\subsection{Description of the study}
Subnetworks are defined on the basis of groups of ROIs (nodes). Once groups of nodes are selected, subnetworks are either connections between nodes of the same group or the connections between two groups of nodes. The choice of decomposition used in the application is not mandatory. Depending on the data, the researcher could use more elaborated decompositions.\\

We used two different prior decompositions of the set of nodes of the global network. The first one corresponds to the decomposition of the brain network into intra/inter lobes subnetworks (LOB). The second is based on a recent study \citep{Chen12} which uses a different approach to cortical localization (CHN). We also decompose the set of nodes into communities (subsets of nodes) using two different data-driven algorithms, Leading EigenVectors (LEV) \citep{Newman2006finding} and Walktrap (WT) \citep{Pons2005computing}. The decomposition methods have been applied only on control data.\\

\subsection{Results}
\begin{figure*}
  \centering
\makebox{\includegraphics[angle=0,width=\textwidth]{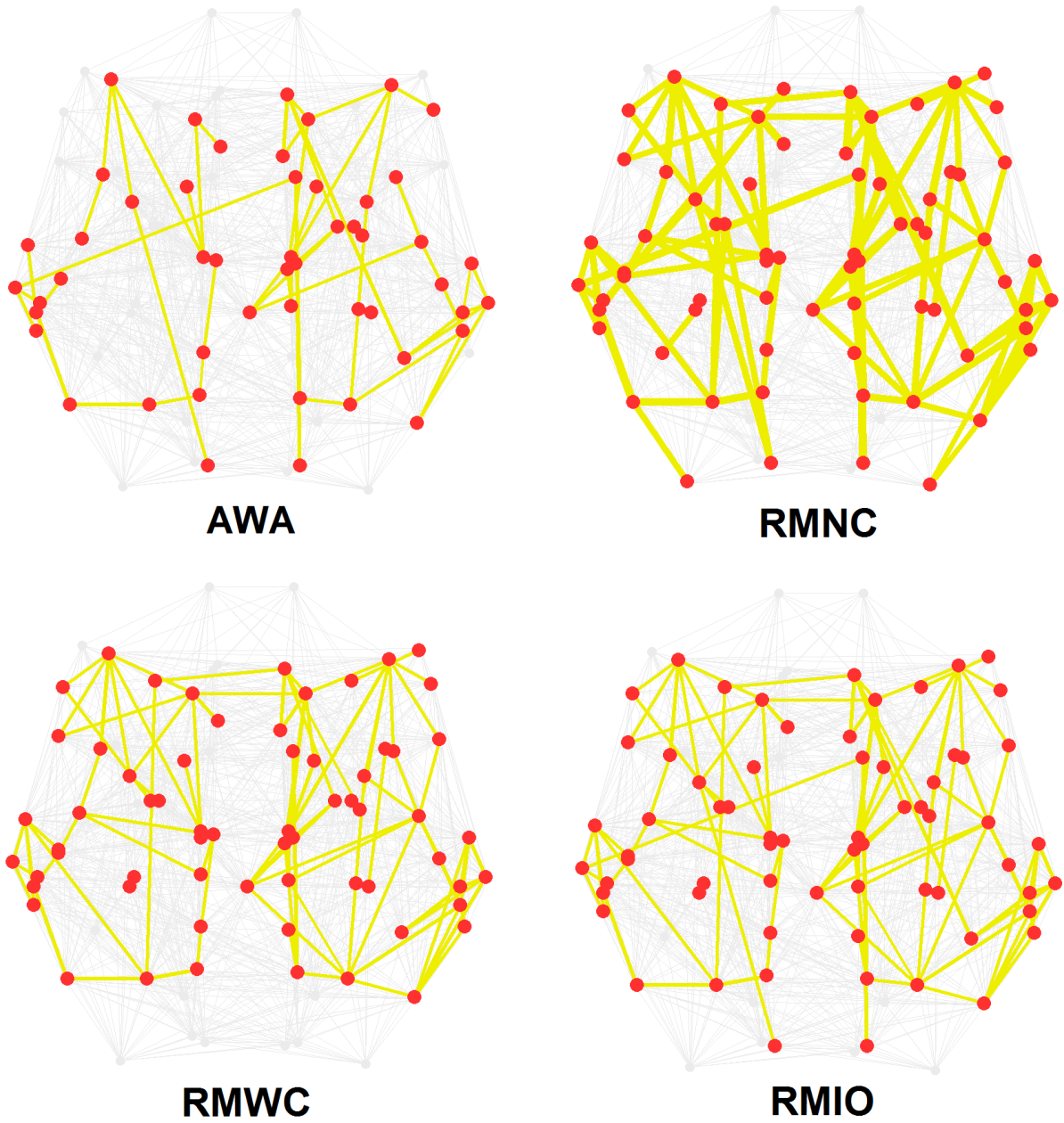}}
\caption{Dorsal view of the human brain network with the significant connections using different methods.}\label{PartIII: Significant connections}
\end{figure*}

We compared the relaxed two-step methods (RMWC, RMNC, RMIO) with the AWA in terms of the number of connections declared to be significantly different. We reported the number of connections/nodes declared to be significant in Table \ref{PracticalExampleResults}. For the relaxed methods, RMNC, RMWC and RMIO we give 5 values. Four values that correspond to the different decompositions: Lobes, Chen, LEV, WT and the fifth value corresponds to the number of common rejections between the four precedent values. This common value highlights the dependence of the relaxed methods on the choice of the decomposition. The results show again the potential gain of the relaxed methods and the relevance in brain connectivity analysis. The results could be improved with an optimal choice of the decomposition especially for the RMWC which has a quite strong dependence on the choice of the decomposition. In Figure \ref{PartIII: Significant connections} we show the significant connections obtained by the compared methods (AWA, RMNC, RMWC an RMIO).\\

\begin{table*}[ht!]
  \centering
  \begin{tabular}{| l | c |c c c c c|c c c c c|c c c c c|}
    \hline \hline
      & \footnotesize{AWA} &  & & \footnotesize{RMWC} & &  &   & & \footnotesize{RMNC} & & & & & \footnotesize{RMIO} & & \\
     &  &\footnotesize{LOB}&\footnotesize{CHN}&\footnotesize{LEV}&\footnotesize{WT}&$\bigcap$ & \footnotesize{LOB} & \footnotesize{CHN} & \footnotesize{LEV} & \footnotesize{WT} & $\bigcap$  & \footnotesize{LOB} & \footnotesize{CHN} & \footnotesize{LEV} & \footnotesize{WT} & $\bigcap$ \\ \hline \hline
    \footnotesize{CW} & 52 & 92 & 88 & 101 & 98 &  81 & 105  & 105  & 101 & 103 & 101 &96&96&101&99&89\\
    \footnotesize{NE} & 38 & 43 & 45 & 50  & 49 &  37 & 40  &  44 & 47 & 44 & 40 &43&46&46&46&42\\
    \footnotesize{NS} & 61 & 67 & 64 & 69  & 69 &  64 &  66 & 66  & 68 & 66 & 66 &66&65&66&67&65\\
    \hline \hline \hline
  \end{tabular}
  \caption{The number of connections/nodes declared to be significant using the different methods AWA, RMWC, RMNC and RMIO. For the relaxed methods, we give 4 values that correspond to the different decompositions: LOB, CHN, LEV and WT. We also report the common number of rejections among the first four values. We note this common value by $\cap$.  Three different network measures are used: connection weight (CW), nodal efficiency (NE) and nodal strength (NS). In both cases, the Bonferroni procedure is used to correct for multiplicity.}\label{PracticalExampleResults}
\end{table*}

\subsection{An extended study}
\begin{figure*}
  \centering
\makebox{\includegraphics[angle=270,width=\textwidth]{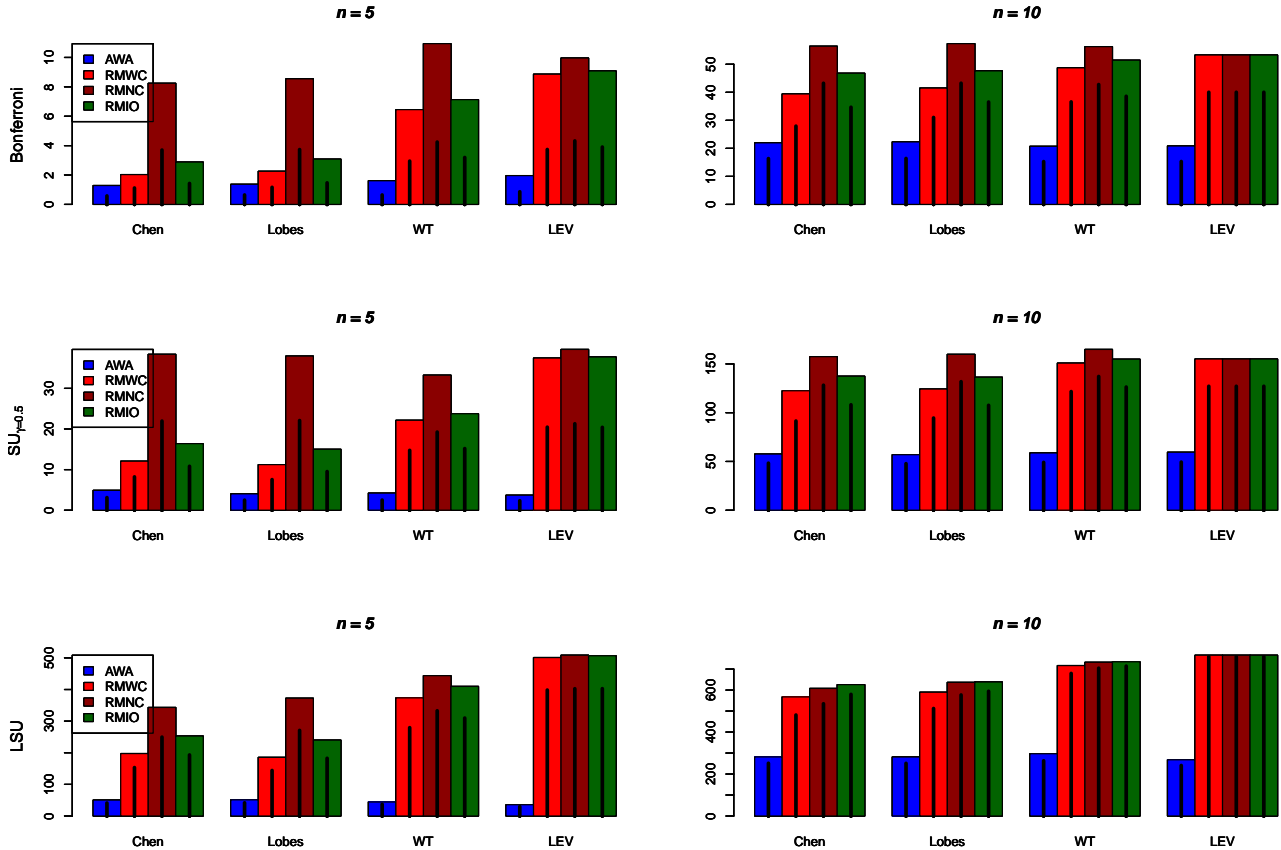}}
\caption[Average number of significant connections using permutations.]{Average number of connections detected as significantly different between the compared groups, using different strategies after 500 iterations. The Bonferroni procedure is used in the first row, the $\SU_{\gamma=0.5}$ in the second row and the LSU procedure in the third row. The number of subjects in each group is either $n=5$ or $n=10$ as indicated. The black bars show the common rejections between the rejections obtained by each method (AWA, RMWC, RMNC or RMIO) and the rejections obtained by the AWA when using the complete sample.}\label{ExtendedResults}
\end{figure*}

Our comparison consists in the following. Among the 16 pre-school children and the 14 adolescent children, we randomly select 5, or 10 subjects from each group, we apply the different multiple comparison procedure (AWA or the two-step procedures) and then, we estimate the number of significant connections. The operation is repeated 500 times. In Figure \ref{ExtendedResults}, we show the average number of significant connections for each strategy, using three different multiple comparison procedures: the Bonferroni procedure, the LSU procedure and the scaled SU procedure with $s(i)=i^\gamma$ and $\gamma=0.5$ as a scaling function. We note this procedure $\SU_{\gamma=0.5}.$ For the relaxed methods, we give two values that correspond to two different decompositions: Lobes and Chen. We also reported the average of the common rejections between the rejections obtained by each method (AWA, RMWC, RMNC or RMIO) and the rejections obtained by the AWA when using the complete sample.\\

Figure \ref{ExtendedResults} clearly shows the potential gain of the relaxed methods and the relevance in brain connectivity analysis. The RMWC should be chosen when we have more confidence on the network decomposition. Otherwise, RMNC is preferable as it has a less strict screening in the first step. This can be seen in the left column of Figure \ref{ExtendedResults} where the number of subjects in each group is 5. However, the relaxation coefficient is much smaller compered to the one obtained with RMWC. The RMIO could be used when the user would like to keep the results obtained by the AWA. We also see in Figure \ref{ExtendedResults} that the $\SU_{\gamma=0.5}$ procedure is intermediate between the Bonferroni procedure and the LSU procedure.\\

\section{Conclusion}
We presented in this study a two-step strategy that exploits the positive dependence between tests without relying on strong assumptions. In the resulting tests, one does not need to estimate the conditional p-values. Only a relaxation coefficient has to be calculated. The two-step procedures can be used to control the $\fwer$, the $\fdr$ or any error rate based on the modified p-values. They perform almost always better than the AWA. Although, the relaxed procedures do not exploit the information of the positive dependence in an optimal way, they are constructed in order to control false discoveries in a wide range of possible behaviors. Nevertheless, the gain reached in the simulations seems to be enough to satisfy the users.
We also presented an adaptation of the two-step methods in a practical example involving real human brain networks. The two-step procedures were applied to brain connectivity analysis and we showed its potential compared to the AWA in terms of significantly different nodes/connections. The relaxed procedures do not exploit the information of the positive dependence in an optimal way.\\

\def\urlprefix{}
\def\url#1{}
\section{Acknowledgments}

This work was supported in part by the FNS grant $N^{0}200020_{-}144467$ and by the Center for Biomedical Imaging (CIBM) of the Geneva-Lausanne Universities and the EPFL, as well as the foundations Leenaards and Louis-Jeantet. Patric Hagmann is supported by the Leenaards Foundation.

\bibliographystyle{chicago}
\bibliography{D:/Thesis/Bib}%

\end{document}